\documentstyle[NATO,epsf]{crckapb}

\def\begineq{\begin{equation}}
\def\endeq{\end{equation}}
\def\ie{{\it i.e.,~}}

\begin{opening}
\title{Velocity Correlations in \protect\\ 
Driven Two-Dimensional Granular Media$^1$}
\author{C. Bizon}
\author{M. D. Shattuck}
\author{J. B. Swift}
\author{Harry L. Swinney}
\institute{Center for Nonlinear Dynamics and Dept. of Physics,\\ University of
Texas, Austin, TX 78712, USA}
\end{opening}

\begin{document}
\begin{abstract}
Simulations of volumetrically forced granular media in two dimensions  produce states with nearly
homogeneous density.  In these states, long-range velocity correlations with
a characteristic vortex structure develop; given sufficient time, the 
correlations fill the entire simulated area.  These velocity correlations
reduce the rate and violence of collisions, so that pressure is smaller for
driven inelastic particles than for undriven elastic particles in the same
thermodynamic state.  As the simulation box size increases, the effects of velocity
correlations on the pressure are enhanced rather than reduced.
\end{abstract}

\vspace{0.3in}

\section{Introduction}
\label{intro}

\footnotetext[1]{Submitted to {\em Dynamics: Models and Kinetic
Methods for Nonequilibrium Many-Body Systems}, Ed. J. Karkheck, Kluwer,
Dordrecht, The Netherlands, Feb. 1999.}
In rapid flows of granular media, the mean time between collisions of grains is much longer
than the duration of a collision~\cite{campbell90};  for such flows, the
machinery of kinetic theory is expected to apply. Continuum 
equations~\cite{lun84,jenkins85} analogous to the Navier-Stokes equations can
be produced, allowing quantitative analysis of flows.   The simplest 
and most common formulations incorporate Boltzmann's assumption of molecular chaos: that particle velocities are uncorrelated.

While this assumption works well for low-density molecular gases, granular 
gases may not abide such a restriction because collisions between grains are 
inelastic. Inelastic collisions reduce relative velocities, so that 
post-collisional velocities are more parallel than pre-collisional velocities.
Repeated inelastic collisions can lead to strong, long-range velocity 
correlations, which standard kinetic theory does not include. We will use
molecular dynamics simulations to produce steady state granular gases and
study the velocity correlations that develop.

The importance and intrinsic interest of velocity correlations in granular
flows have been noted by a number of researchers.  Two-dimensional simulations of an initially
homogeneous distribution of inelastic disks without velocity correlations show that
as time progresses, velocity correlations build in both strength and range~\cite{orza98}.  These simulations are limited in time, however, because the 
homogeneous state is unstable to density fluctuations, and rapidly becomes
inhomogeneous. Nevertheless, these simulations clearly displayed a 
characteristic vortex structure of the correlations. Based upon similar considerations, ring kinetic theory, which
accounts for velocity correlations, has been applied to the cooling 
state~\cite{vannoije98}.  One-dimensional simulations of stochastically forced 
point particles also show velocity correlations~\cite{swift98}. 

We apply stochastic forcing~\cite{swift98,williams96} to two-dimensional 
event-driven simulations of inelastic disks.  The forcing overcomes the
tendency of the granular material to form density clusters, and approximately
homogeneous steady states form.  In an earlier study of these 
states~\cite{bizon99}, we found strong velocity correlations that extended
throughout the entire simulation area. In the present work, we discuss
the simulation method, show that the velocity correlations are essentially
independent of the simulated area, and describe the vortex structure of the
correlations.

\section{Simulations of Driven Granular Gases}

We treat collisions between molecules as instantaneous and binary.  
The collisions between grains conserve momentum but dissipate energy.  
Between collisions, particles travel along straight lines if unaccelerated,
or along parabolas if accelerated.  This model allows efficient simulation
of collections of particles using event-driven molecular dynamics~\cite{lubachevsky91,marin93}.

When particles collide, the component of the relative particle velocity along the line joining particle centers, $v_n$, is reversed, and reduced by a factor
$e$, the coefficient of restitution, which can take values between 1 for elastic particles and 0 for completely inelastic particles.  We allow $e$ to depend 
on $v_n$ through
\begineq
e(v_n) = \left \{\begin{array}{cc} 1 - B v_n ^ {\beta} &, v_n <
v_o \\ \epsilon &, v_n > v_o
\end{array}
\right.  ,
\label{restform}
\endeq
where $B = (1-\epsilon)(v_o)^{-\beta}$, $\beta=3/4$ and $\epsilon$ is a constant, chosen to be $0.7$.   These parameters give quantitative
agreement to experiments on patterns in vertically oscillated granular media~\cite{bizon98,debruyn98}.  The variation in $e$ has the effect of removing
inelastic collapse~\cite{goldman98}, which is a singularity in the 
inelastic hard sphere model that produces an infinite number of collisions within a finite time~\cite{mcnamara92,mcnamara94}.  In general, 
colliding particles also exert frictional forces on one another; for this
paper, we assume that the coefficient of friction is zero, so that we are studying only the effects of inelasticity.

Because of inelasticity, the energy of an unforced collection of grains
inevitably decreases.  To achieve steady states, then, we must force the
granular material.  Methods that force through boundaries, such as shaking,
invariably produce strong inhomogeneities in the system; to achieve 
near-homogeneity, we force volumetrically, assuming the particles to
be in contact with a white-noise heat bath~\cite{williams96}.  Whenever
two particles collide, the velocities of two other randomly selected particles 
are changed by amounts $|\delta {\bf v}| \hat{\bf r}_i$, where the magnitude of the kicks, $|\delta {\bf v}|$, are always the same, but the direction vectors, 
$\hat{\bf r}_i$ are randomly chosen for each kicked particle.  In addition
to the white noise heat bath, we perform a lesser number of runs with two
other heat baths.  To model the motions of pucks on an air 
table~\cite{oger96,ippolito95}, we can allow particles to accelerate randomly 
from collision to collision.  Finally, we model the effects of a strong heat
bath, which we denote the Boltzmann bath, by completely obliterating the
velocities of randomly chosen particles, and giving new velocities based on 
a Boltzmann distribution.  The details of all three forcing methods may be found in ~\cite{bizon99}.

We perform simulations of N disks of diameter $\sigma$ moving in a 
two-dimension square of side length $L$, which varies from $52.6 \sigma$ to
$420.8 \sigma$.  The simulation box is periodic in both directions.  The 
solid fraction, defined as $N {{\pi}\over{4}} {{\sigma^2}\over{L^2}}$, is
$0.5$ for all runs.  Because of the variation of $e$ with relative normal
velocity, the velocity scale $v_0$ enters; we use $v_0$ to nondimensionalize
velocities, and $v_0^2$ to nondimensionalize the granular temperature $T$.
For $T$ much larger than one, most particle collisions will occur with the
high-velocity value of $e$, $0.7$; for lower $T$, a range of $e$ will occur.

\section{Dependence of Correlations upon Simulation Area}

We denote two particles $1$ and $2$, and ${\bf \hat{k}}$ the a unit vector
pointing from the center of $1$ to the center of $2$.  The velocity of $1$
then has a components parallel to, $v_{1}^{||}$, and perpendicular to, $v_{1}^{\perp}$, ${\bf \hat{k}}$, as does particle $2$.  We define two correlation functions
\begin{eqnarray}
\langle v_{1}^{||} v_{2}^{||} \rangle &=& \sum v_{1}^{||} v_{2}^{||} / N_r,\\
\langle v_{1}^{\perp} v_{2}^{\perp} \rangle &=& \sum v_{1}^{\perp} v_{2}^{\perp} / N_r,
\end{eqnarray}
where the sums are over the $N_r$ particles such that the distance between
the two particles is within $\delta r$ of $r$.  For uncorrelated particle
velocities, $\langle v_{1}^{||} v_{2}^{||} \rangle$ and $\langle v_{1}^{\perp}
v_{2}^{\perp} \rangle$ will both give zero.

\begin{figure}
\epsfxsize=.95\textwidth
\centerline{\epsffile{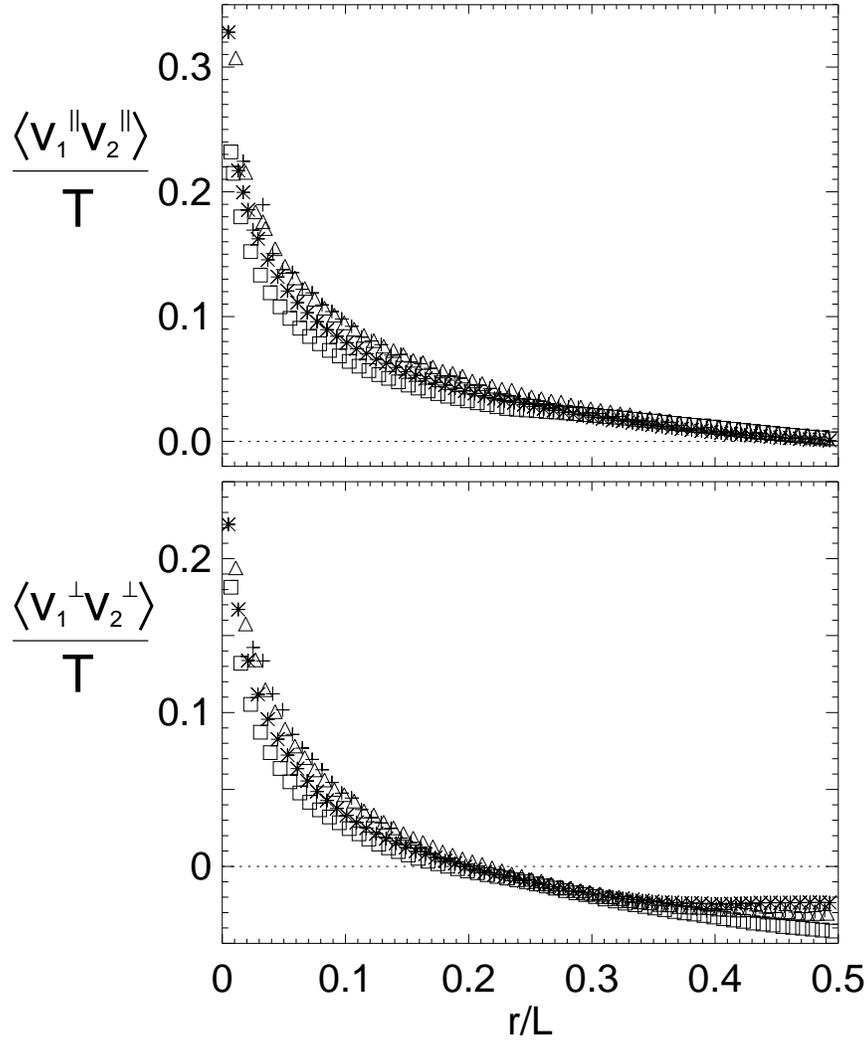}}
\smallskip
\caption{Velocity
correlations as a function of particle separation at $\nu=0.5$ and $T \approx
30$, for four different box sizes. $+: L=52 \sigma$, $\triangle: L=105 \sigma$,
$*: L=211 \sigma$, $\Box: L=421 \sigma$.} 
\label{vcorrsize}
\end{figure}

In the smallest simulation area, $L=52.6\sigma$, correlations extend the full
length of the computational cell.
Cell filling structures may be divided into two cases:   structures with a
natural length that is larger than the box in which they exist and structures
that will always grow to fill any finite box.   To differentiate between the
former and the latter, we performed four simulations with white noise forcing,
quadrupling the area at
each step, while holding the solid fraction fixed at $0.5$. The granular temperature $T$ is 
approximately $30$, but varies between $28$ in the smallest box and $32$ in the largest. This variation in temperature is not important; for $T>>1$, the coefficient of restitution is independent of collision velocity. In this limit, the role of the temperature is simply to set the velocity scale.
The velocity
correlation functions are shown in Fig.~\ref{vcorrsize}.  Even in the largest
simulation, composed of 112768 particles,  the correlations fill the box.
However, the correlation functions for the largest simulation are somewhat
different from the smaller ones.  This is probably due to poorer statistics; in
terms of collisions per particle, this run lasted only one-half as long as the
next largest.

Because velocity correlations are positive for small separations, particles
collide less frequently and with less relative velocity than elastic particles
at the same density, for which velocity correlations are much smaller.  As
a result, less momentum will be transferred through inelastic collisions than
through elastic collisions, and the pressure, $P$, will decrease.

Assuming that velocity correlations do not exist, the equation of state for
dense granular gases is given by~\cite{jenkins85}
\begineq
P = (4/\pi\sigma^2) \nu T (1 + (1+e) G(\nu)).
\label{state},
\endeq
The first term on the right hand side, $(4/\pi\sigma^2) \nu T$, accounts for momentum transfer due to 
particle streaming without collisions, while the second term, $(4/\pi\sigma^2) \nu T (1+e) G(\nu)$, accounts for
the momentum transfer due to particle collisions~\cite{chapman}.  In the absence of velocity
correlations, $G(\nu)$ is defined as $\nu g(\nu,\sigma)$, where $g(\nu,\sigma)$
is the radial distribution function for the particles, evaluated at 
zero particle separation.  Calculation of $P$ from simulation, via 
measurement of the virial~\cite{rapaportsbook}, becomes a measurement
of $G(\nu)$, which describes the collisional momentum transport. 
If velocity correlations exist, $G(\nu)$ will be reduced, since less
momentum will be transported collisionally.

\begin{figure}
\epsfxsize=.9\textwidth
\centerline{\epsffile{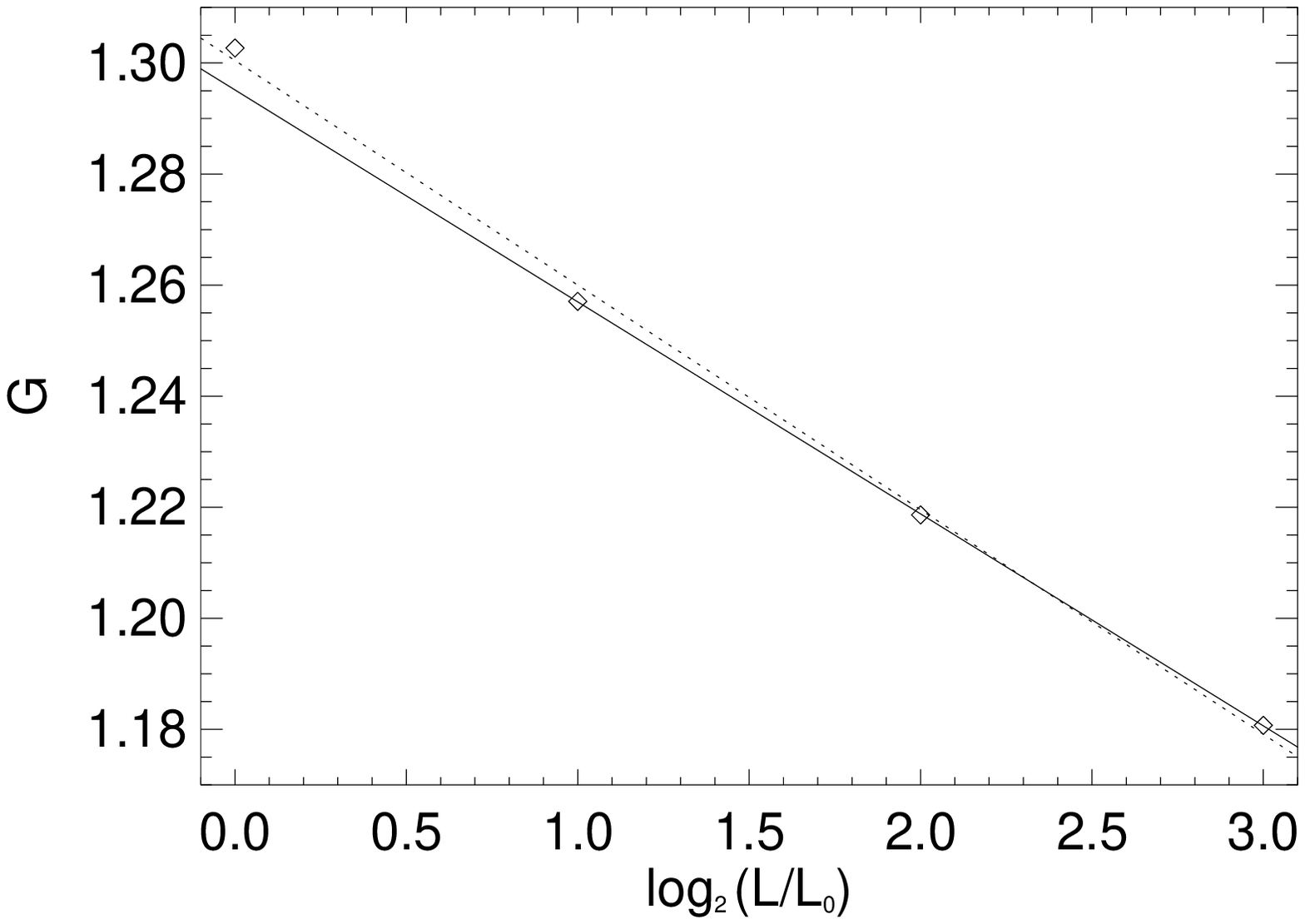}}
\smallskip
\caption{$G$ as a function of $L$ for the
runs shown in Fig.~\protect{\ref{vcorrsize}}. $L_o=52\sigma$ denotes the
length of the smallest box.  The dotted line is a fit to all four points: $G(\nu) = 1.3 - 0.04 \log_2(L/L_o)$, The solid line as a fit to the three largest $L$
values: $G(\nu) = 1.295 - 0.038 \log_2(L/L_o)$. Note that the log is base 2.}
\label{GvsL}
\end{figure}

Figure ~\ref{vcorrsize} shows that the short range velocity correlations
depend on the size of the box; therefore, $G(\nu)$ should also depend on $L$.
Figure~\ref{GvsL} displays $G(\nu)$ as a function of $L$ for these four runs.
Over about one decade, $G(\nu)$ scales with $\log{L}$.  Clearly this scaling
can not continue indefinitely, since unphysical negative values of $G(\nu)$ 
would result. Note also, that increasing the box size actually leads to 
values of $G(\nu)$ farther from the values for uncorrelated velocities.

\begin{figure}
\epsfxsize=.9\textwidth
\centerline{\epsffile{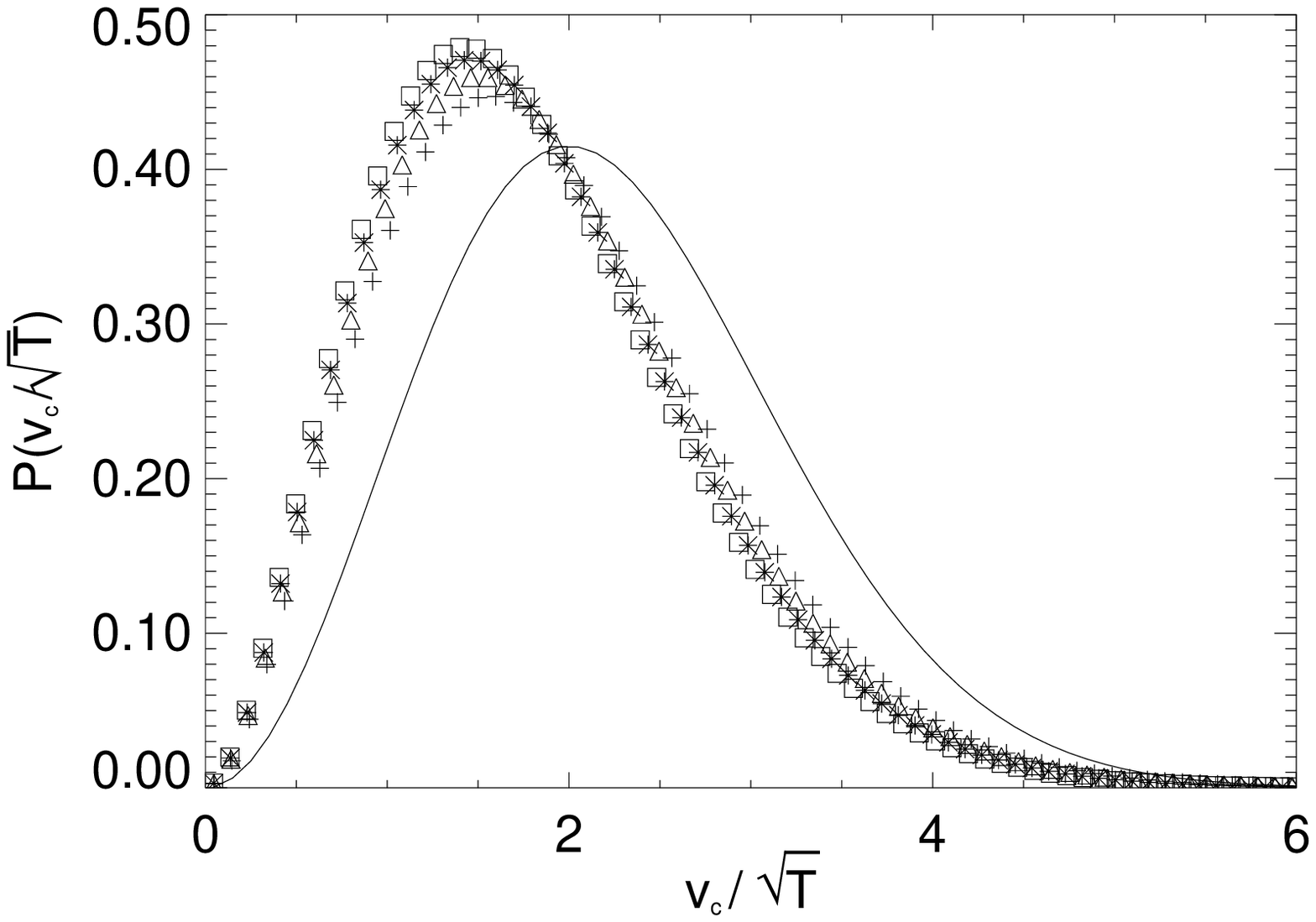}}
\smallskip
\caption{Probability distribution of collision velocities $v_c = |{\bf v}_1 - {\bf v}_2|$, for the data in Figs.~\protect{\ref{vcorrsize}} 
and~\protect{\ref{GvsL}}.  
$+: L=52 \sigma$, $\triangle: L=105 \sigma$,
$*: L=211 \sigma$, $\Box: L=421 \sigma$. The solid curve is $P(v_c/\sqrt{T}) =  (1/2\sqrt{\pi T^3}) v_c^2 e^{-v_c^2/4T}$, which holds for elastic particles.}
\label{PvsL}
\end{figure}

This unusual result, that the importance of velocity correlations increases
with increasing computational area, can also be deduced from the 
distribution of collision velocities.  Figure~\ref{PvsL} exhibits these
distributions for the runs displayed in Figures~\ref{vcorrsize} 
and~\ref{GvsL}.  As the computational area increases, so too does the 
deviation from the distribution predicted for particles chosen without
correlation from a Boltzmann distribution, plotted as a solid curve.

\section{Vortex Structure}

Inelasticity breeds velocity correlations; reduction of relative velocity
in collisions leads to particles moving more alike after collisions than
before.  On average, then, particles will be surrounded by particles that
are moving along with them.  The structure of the velocity correlations can
be elucidated by calculating this average flow around each particle.

For a single particle $i$, we can calculate the flow around it by translating
it to the origin, and rotating so that its velocity lies along the positive
$x$ axis.  If ${\bf v}(x,y)$ is the velocity field defined by the particles,
then the flow around particle $i$ is given by
\begineq
{\bf u}_i = R_{\theta(i)} {\bf v}(x-x_i,y-y_i),
\label{floweqn}
\endeq
where $\theta(i)$ is the angle between the $i$-th particle velocity, ${\bf v}_i$
, and the positive $x$ axis, $(x_i,y_i)$ is the position of the $i$-th particle,
 and $R_\theta$ is the  operator
that rotates vectors clockwise through angle $\theta$.
The average flow around particles, then, is
\begineq
{\bf u} = \sum_{i=1}^N {\bf u}_i / N.
\endeq
Finally, ${\bf u}$ is averaged over about 100 frames to reduce noise.

\begin{figure}
\epsfxsize=.9\textwidth
\centerline{\epsffile{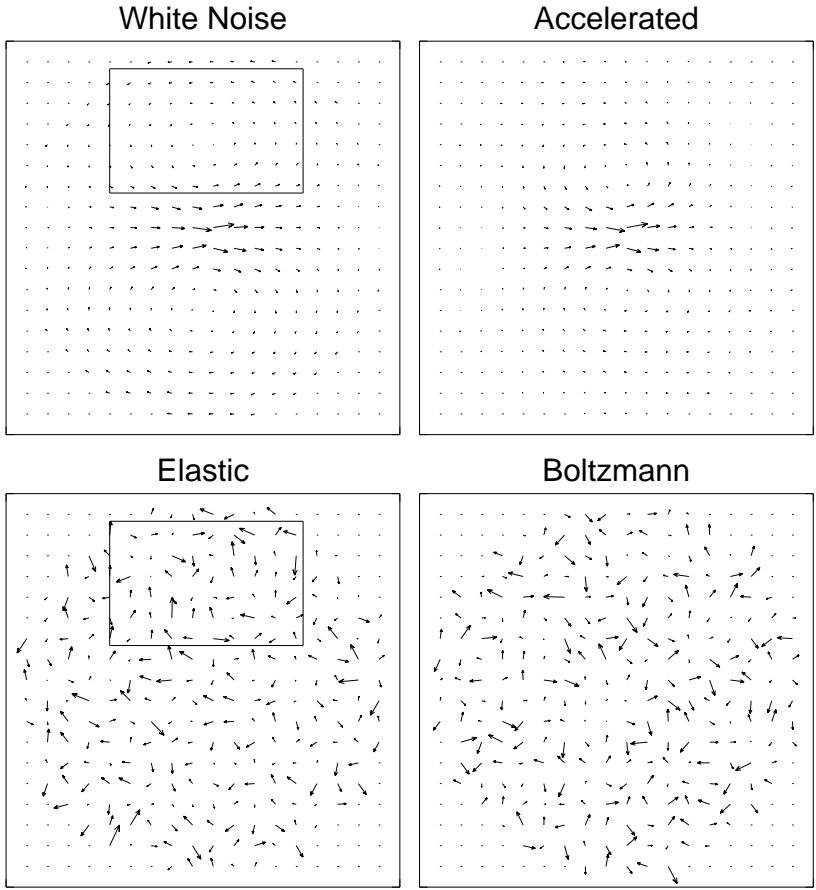}}
\smallskip
\caption{The average velocity fields around a particle centered in
each cell and moving to the right, ${\bf u}$, for elastic particles
and for inelastic particles forced in three different ways (cf section
2).  Each vector field is scaled separately so that its
longest vector has length one.  Compared to the (suppressed) central vector,
these lengths are: White noise, 0.2; Accelerated, 0.27; Boltzmann, 0.008; Elastic 0.008. The boxed regions in the white noise and elastic flows are shown in
 Fig.~\protect{\ref{closeup}}.}
\label{Flow}
\end{figure}

Figure ~\ref{Flow} displays vector fields of the  average flow around
particles, ${\bf u}$, for the three types of forcing, as well as for unforced
elastic particles, all at $\nu=0.5$ and $T=1.05$. In each case, the vector
at the origin, which measures only the average particle speed, has been
suppressed, and the longest remaining vector in each field has been scaled to
unit length.  In both the white noise and accelerated forcings, the average
flow near the origin is along the positive $x$ axis, \ie with the direction
of the central particle's motion.  The Boltzmann bath shows some indications
of this effect close to the origin, but the correlations are destroyed by
the strongly thermalizing forcing before they can propagate to larger length
scale.  For the elastic particles, there is no discernible flow, only noise.

\begin{figure}
\epsfxsize=.9\textwidth
\centerline{\epsffile{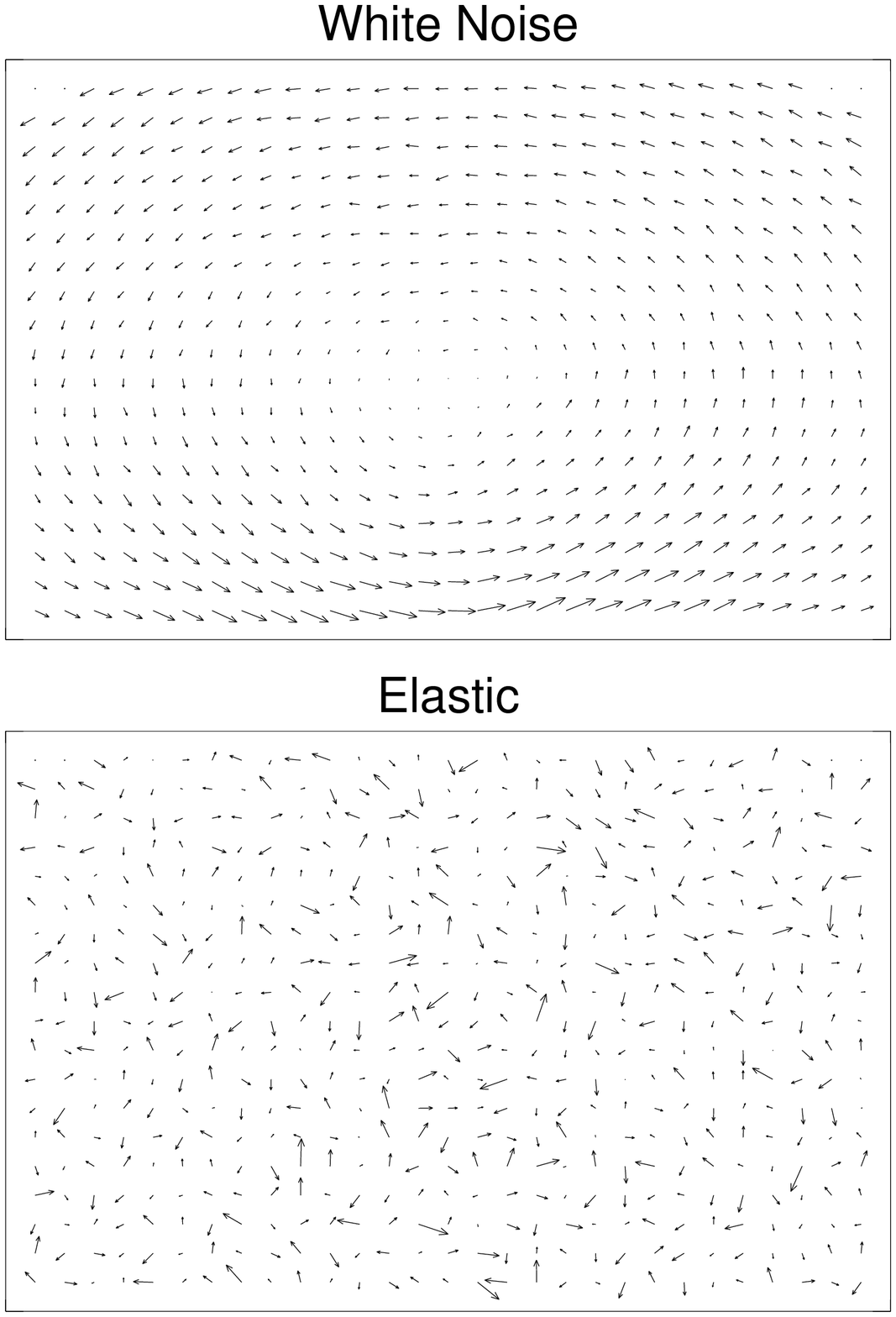}}
\smallskip
\caption{A close-up on the boxed regions in
Fig~\protect{\ref{Flow}} reveals that for inelastic particles, large vortices
form, one on each side of the particle. The longest vector in the velocity
 field for inelastic particles represents a velocity nine times larger than
that represented by the longest vector for elastic particles.}
\label{closeup}
\end{figure}

Close to any particle, surrounding particles move along with it.  Farther away,
the
correlations decay and cannot be seen on Fig~\ref{Flow}, so the boxed
regions for the white noise forcing and for elastic particles are expanded
in Fig.~\ref{closeup}.  While expansion of the velocity field for
elastic particles produces
still more noise, the inelastic flow field reveals a highly ordered vortex
structure.   Along the direction of the central particle's motion, the velocities slowly drop to zero, while perpendicular to the original particle's motion,
the velocities drop to zero and increase in the negative direction;  this flow
makes clear the structure of the velocity correlation functions in
Fig.~\ref{vcorrsize}.

This vortical flow is reminiscent of similar structures produced
in simulations of elastic particles~\cite{alder68,alder70a} by Alder and
Wainwright.  In their simulations, they discovered diffusive behavior different
from that predicted by kinetic theory.  The diffusion constant may be written
in terms of the slope of the exponentially decaying autocorrelation function.
However, Alder and Wainwright found deviations from exponential decay, and
traced the deviations to a vortical flow.  If particles $a$ and $b$ are initially
uncorrelated, an elastic collision will correlate each particle's  post
collision velocity with the other particle's pre-collision velocity; both
particles now have a correlation with the original velocity of particle $a$.
As particle $b$ collides with other particles, they gain information
about particle $a$'s initial velocity.   Several collision times later,
this information has been transmitted to many particles.

There are two main differences between the vortices in flows of elastic particles and those in flows of inelastic particles.
Alder and Wainwright produced the flow field given by
\begineq
{\bf u}(t')_i = R_{\theta(i,t)} {\bf v}(x-x_i(t),y-y_i(t),t').
\endeq
For $t'=t$, Eq.~\ref{floweqn} is recovered; for elastic particles, no
structure is apparent.  It is only at later times, $t' > t$, that a vortex 
appears in
${\bf u}(t')$.  For the inelastic particles, however, structure is clear
at $t'=t$.  The second difference is the strength of the vortex.  The
strongest velocity in Alder and Wainwright's vortex was about 2\% of the
original velocity, while for inelastic particles, the strongest velocity
can be about 40\% of the central velocity.

\begin{figure}
\epsfxsize=.9\textwidth
\centerline{\epsffile{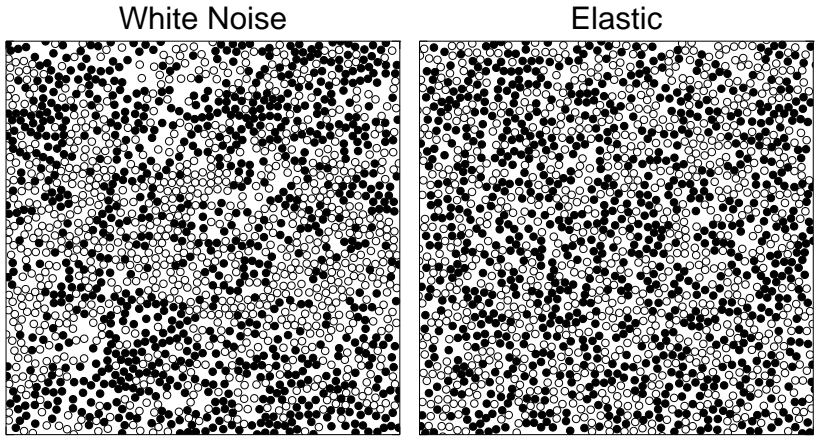}}
\smallskip
\caption{Snapshots of simulations with white noise driving
and with elastic particles; large coherent structures are visible for the
dissipative system on the left.  Particles
with positive horizontal velocity are black, particles with negative horizontal
velocity are white. ($\nu = 0.5, T=1.05$)}
\label{ParticleFields}
\end{figure}

The inelastic vortex is so strong that hints of it are visible
even in a single snapshot of particles.  Figure~\ref{ParticleFields} shows
such a snapshot, with particles colored black if they have positive
horizontal velocity and white if they have a negative horizontal velocity.
For elastic particles, the black and white are well mixed, but in the
inelastic case larger scale structure can be glimpsed.  Black particles
are concentrated along the top and bottom of the image, and white particles
are concentrated along the central region.

\section{Conclusion}

The correlations we have found are consistent with those of simulations on the
homogeneous cooling state~\cite{orza98}.  In those simulations, the range of
velocity correlations grew until the onset of large scale density variations.
The addition of forcing in our simulations suppresses the growth of
density fluctuations, allowing the velocity correlations to continue to grow
until they extend throughout the entire computational area.

The results we obtain are not particularly sensitive to the exact form
of the forcing.  In both the white noise and accelerated forcing schemes, 
vortical correlation structures form.  Only when the bath explicitly 
destroys correlations, as in the Boltzmann bath, do the results differ. 

\section{Acknowledgments}

We thank J. T. Jenkins, M. H. Ernst, and T. P. C. van Noije for useful 
discussions.  This work was supported by the Department of Energy Office of
Basic Energy Sciences.

\bibliography{/mich2/bizon/grain/bib/sand}
\bibliographystyle{myprsty}

\end{document}